\documentclass[aps,prl,twocolumn,groupedaddress,showpacs]{revtex4}

\begin{document}

\title{Weak measurements with orbital angular momentum pointer states}

\author{G. Puentes$^{1}$}
\email[]{graciana.puentes@icfo.es}

\author{N. Hermosa$^{1}$}

\author{J. P. Torres$^{1,2}$}
\affiliation{$^{1}$ICFO - Institut de Ciencies Fotoniques,
Mediterranean Technology Park,
08860 Castelldefels, Barcelona, Spain\\
$^{2}$Department Signal Theory and Communications, Campus Nord D3,
Universitat Politecnica de Catalunya, 08034 Barcelona, Spain }


\date{\today}

\begin{abstract}
\noindent Weak measurements are a unique tool for accessing
information about weakly interacting quantum systems with minimal
back action. Joint weak measurements of single-particle operators
with pointer states characterized by a two-dimensional Gaussian
distribution can provide, in turn, key information about quantum
correlations which can be of relevance for quantum information
applications. Here we demonstrate that by employing
two-dimensional pointer states endowed with orbital angular
momentum (OAM), it is possible to extract weak values of the higher
order moments of single-particle operators, an inaccessible quantity with
Gaussian pointer states only. We provide a specific example that illustrates the advantages of our method both, in terms of 
signal enhancement, and information retrieval.
\end{abstract}

\pacs{03.65.Ta, 42.50.Xa, 03.65.-w}

\maketitle

Ideal von Neumann measurements in quantum mechanics are able to
distinguish between the different eigenstates of a given
observable. Contrary to this, weak measurements, a concept first
introduced by Aharonov, Albert, and Vaidman \cite{AAV}, describe a
situation where the coupling between the measuring device and the
observable is a weak perturbation. In this case, the uncertainty
in the measurement is large in comparison with the separation
between the eigenvalues of the observable, such that the different
outcomes can not be resolved.

What makes a weak measurement an interesting phenomenon is that
the weak value of an observable $A$ of the system can yield an
unexpected result \cite{duck1989,WM1,WM2}. In particular, given a
preselected state $|i\rangle$ and a postselected state $|f\rangle$ of the
system, the weak value $\langle A \rangle_{w}$ is defined as:
\begin{equation}
\label{Eq1} \langle A \rangle _{w}=\frac{\langle f |A
|i\rangle}{\langle f |i \rangle}.
\end{equation}
If an appropriate postselection is made, for instance by choosing
$|i \rangle$ and $|f \rangle$ to be nearly orthogonal, the weak
value $\langle A \rangle_{w}$ can be significantly outside the
spectrum of $A$ (weak-value amplification), and it can even take imaginary values.

In the case of an observable $A$, which is weakly coupled to a one-dimensional pointer state, the Hamiltonian of the weakly coupled
system can be described by $H=gAP_{x}$, where $P_{x}$ is the
pointer momentum operator, conjugate to its position operator $X$.
For a sufficiently weak coupling strength $gt$, the interaction
shifts the mean position of the pointer by an amount $\Delta x=gt
\Re(\langle A \rangle_{w})$, where $t$ is the duration of the
interaction. By measuring the mean value of the pointer position
displacement $\langle X \rangle$ and the pointer momentum
displacement $\langle P_{x} \rangle$, it is possible to obtain
either the real part of the weak value $\Re (\langle
A\rangle_{w})$ or its imaginary part $\Im(\langle A \rangle_{w})$
\cite{AS}, thus providing a full measurement of the weak value of
the observable $\langle A \rangle _{w}$.

Several successful experimental implementations of weak
measurements have been accomplished up to date in a wide range of
scenarios: for demonstrating wave-particle duality in the context
of cavity-QED \cite{foster2000,wiseman2002}; for characterizing
the response function of a highly dispersive system
\cite{solli2004}; for the realization of Leggett-Garg inequality violations \cite{Leggett-Garg}; for detecting tiny spatial shifts
\cite{ritchie1991, hosten2008} and tiny temporal shifts \cite{brunner2010}, or tiny beam deflections
\cite{dixon2009}, of intense optical beams; for the direct measurement of 
the wave function of a quantum system \cite{lundeen2011}, or the
weak-value of the polarization degrees of freedom of entangled photon
pairs \cite{pryde2005}.

In many applications, it can be important to access the mean value
of products of single-particle operators, i.e.,  $\langle AB
\rangle$.~Such joint measurements are of great relevance for
quantum information, since they can contain information about
quantum correlations (entanglement) between different degrees of
freedom, as for instance in cluster-state quantum computation
\cite{Briegel}. However, a strong measurement of joint mean values
requires a non-linear Hamiltonian which in many cases can prove
hard to engineer \cite{Turchette}.~Resch and Steinberg \cite{AS},
circumvented this limitation  by employing a two-dimensional Gaussian pointer state, and a
weakly coupling linear Hamiltonian of the form $H=g_{A}AP_{x} +
g_{B}BP_{y}$, where $(P_{x},P_{y})$ are the pointer momentum
operators, conjugate to the pointer position operators $(X,Y)$. By 
performing a second order expansion in the two-dimensional pointer
displacement $\langle XY \rangle$, the authors showed that it is possible to extract the real part of the joint
weak value $\langle AB \rangle_{w}$, for the case of commuting observables $[A,B]=0$.

~In this Letter, we show that by employing pointer states endowed with
orbital angular momentum (OAM), it is possible to retrieve a wider range of
second order weak values.~An important limitation of Gaussian
pointer states is that, due to symmetry properties, they can not
provide access to weak values of higher order moments of single-particle
operators, such as $\langle A^n \rangle_{w}$ or $\langle B^n
\rangle_{w}$, or to higher order moments of joint operators, such
as $\langle A^{n}B \rangle_{w}$ or $\langle B^{n}A \rangle_{w}$
($n>1$), for the case $[A,B]=0$ \cite{AS}. Pointer states with OAM introduce a different
symmetry for the expectation values of operators, and thus provide
access to  the weak values of higher order moments of
single-particle operators and joint operators, a result which is
not attainable with Gaussian pointer states.

Moreover, in many applications employing Gaussian pointer states, the weak amplification factor can only be attained in the imaginary component of the weak value.~This can be seen as an advantage, as the amplified imaginary part can have a simpler operational interpretation than its real counterpart, since it is not tied to the conditioned average value of an observable, and is rather linked to the measurement back action \cite{AS,Dressel}. However, in order to extract such imaginary weak values, additional measurements on the pointers conjugate variable are required. Here we will show that OAM pointer states can
provide access to the enhanced imaginary part  of
higher order weak values, thus outperforming  Gaussian
pointer states in terms of signal enhancement, and information retrieval. 

Finally, we note that since the analytical determination of upper bounds on amplification under the assumption of Gaussian pointer wave-function \cite{Upperbound}, much effort has been devoted to the engineering of optimal probe states  \cite{Shikano}. Here, we present a realistic, and experimentally feasible, pointer distribution that can provide significant advantages over Gaussian probe states, and which could find several applications both in the context of quantum information and foundational tests of quantum mechanics  \cite{Bell}, as well as in weak measurements of cosmological effects, such as in gravitational wave-detection with higher order Laguerre-Gauss beams \cite{Granata}, or the back action of the Hawking radiation from a black hole \cite{Hawking}. \\

Consider the weak interaction between two observables, $A$ and $B$, of a
single-particle system initially prepared in the state
$|i\rangle$, with a measuring device initially prepared in the
state $|\phi\rangle=\int dx dy \phi(x,y)|x,y\rangle$. The total
input state is $|\psi^{in}\rangle=|i\rangle \otimes |\phi\rangle$.
The Hamiltonian of weak interaction, $H=g_A A P_x+g_B B P_y$, describes
the coupling of observable $A$ with the $x$-dimension of the
measuring device, while observable $B$ is coupled to its $y$-dimension.~Here $x$ and $y$ are cartesian coordinates, and their associated  position and momentum quantum operators satisfy the usual commutation rules $[X,Y]=0$ and $[P_{x},P_{y}]=0$, respectively. Also, we restrict to the case of commuting observables of the form $[A,B]= 0$, though more complex expressions can also be obtained in the case where $[A,B] \ne 0$.

We are interested in the mean value of the operator
$O_{XY}=|f\rangle \langle f|XY$ at time $t$,~where the projector $|f\rangle \langle f|$ performs the postselection operation on the system, and commutes with all the spatial observables for the measuring device. The time-dependent mean value $\langle O_{XY}(t) \rangle$ can be
obtained by expanding the Heisenberg's equation of motion to
second order in the coupling parameters $g_A$ and $g_B$ \cite{AS}:
\begin{equation}
\label{Eq2} \langle O_{XY} (t)\rangle = \langle O_{XY} (0)\rangle
+ \frac{it}{\hbar}\langle [H,O_{XY}]\rangle -
\frac{t^2}{2\hbar^2}\langle[H,[H,O_{XY}]]\rangle.
\end{equation}

The two-dimensional pointer states considered here are described
by Laguerre-Gauss (LG) modes. LG modes are a set of solutions of the paraxial wave-equation
\cite{Siegman}, characterized by two integer indices, $p$ and $l$.
The index $p$ is a positive integer, and $p+1$ determines the
number of zeroes of the field along the radial direction. The
winding index $l$, which can take any integer number, determines the
azimuthal phase dependence of the mode, which is of the form $\sim
\exp \left( i l \varphi\right)$. Each mode carries a well-defined
orbital angular momentum of
$l\hbar$ per photon, associated with their spiral wave fronts  \cite{allen1}. The OAM states of light allow
for a relatively simple experimental generation, filtering,
detection, and control \cite{torres_book}. In this paper, we will
concentrate on the case $p=0$ and $l=0,\pm 1$, since this is enough to
demonstrate the benefits of using OAM pointer states.

 Specifically, the two-dimensional pointer distribution considered here is described by \cite{Padgett}:

\begin{equation}
\label{Eq5} \phi(x,y)=N \left( x + i \text{sgn}(l) y
\right)^{|l|} \exp \left(-\frac{x^2+y^2}{4\sigma^2}\right), 
\end{equation}

where $\sigma$ is the uncertainty in the pointer state, and $N$ is
a normalizing constant so that $\int dx dy |\phi(x,y)|^2=1$.  The
case $l=0$ corresponds to a pointer state with a 2D-Gaussian
distribution. In this case, the pointer state is factorable in the
two directions, and therefore can not be used to retrieve
higher order weak moments of $A$ and $B$. The case $l=\pm 1$
corresponds to states endowed
with orbital angular momentum (OAM). Now the pointer distribution
is no longer factorable, and this is a key factor to retrieve
higher order weak moments of the form $\langle A^2 \rangle_{w}$
and $\langle B^2 \rangle_{w}$, as it will be shown below. Moreover, we note that by considering larger values of $l$, it
should be possible to access weak values of a wider range of
moments of single-particle operators and joint operators.\\

Inspection of Eq. (\ref{Eq2}), and symmetry properties of Gaussian
integrals related to the specific shape of the pointer states
given by Eq. (\ref{Eq5}), show that the first non-zero term is
$\langle [H,[H,O_{XY}]] \rangle$, which is second order in the
coupling parameters. Higher order terms  do not vanish either, but
they can be considered negligible if the coupling constants $g_A$
and $g_B$, and the duration of the interaction $t$, are
sufficiently small with respect to the pointer uncertainty $\sigma$.~By
making use of $\langle \phi |P_{x}X| \phi \rangle=\langle \phi
|P_{y}Y| \phi \rangle  = -i\hbar/2$, and $\langle \phi |XP_{x}|
\phi \rangle=\langle \phi| YP_{y}| \phi \rangle  = i \hbar/2$, and
due to symmetry properties of the chosen pointer states, we obtain:

\begin{eqnarray}
\label{Eq6} \langle XY \rangle &= &  \frac{g_{A}g_{B}t^2}{2}\left[
\Re(\langle AB\rangle_{w})  +
\Re(\langle A\rangle_{w}^{*} \langle B\rangle_{w}) \right]  \nonumber \\
& & + l \frac{t^2}{2}\left[ g_{A}^2 \Im(\langle A^2 \rangle_{w}) +
g_{B}^2 \Im(\langle B^2 \rangle_{w})\right].
\end{eqnarray}

For $l=0$ (Gaussian pointer states),
we recover the result given in Ref.~\cite{AS}.~For $l=\pm 1$ (OAM pointer states), we obtain additional terms
proportional to $g_{A}^2 t^2 \Im(\langle A^2
\rangle_{w})$ and $g_{B}^2 t^2 \Im(\langle B^2 \rangle_{w})$, thus
providing access to the imaginary part of the second order weak
values, a quantity which can not be retrieved using Gaussian
pointer states only.\\

Note that in an experiment one measures $\langle XY \rangle$,
which contains both the joint weak value $\langle AB \rangle_{w}$,
and the squared weak values of single-particle operators $\langle
A^2\rangle_{w}$ and $ \langle B^2 \rangle_{w}$.~In order to
address the second order weak moments of $A$ and $B$, we can perform
two separate measurements, a first one with a Gaussian pointer
state, thus obtaining Eq. (\ref{Eq6}) with $l=0$,~and a second
one with an OAM pointer state. By subtracting these two
measurements, it is possible to obtain the imaginary part of the
second order weak value. Also, by measuring the combined position-momentum pointer shift  $\langle YP_{x} \rangle$, it is possible to access a term proportional to  $\Re( \langle A^{2}\rangle_{w}) $, while by measuring the complementary  shift
$\langle XP_{y} \rangle$, one can access a term proportional to $\Re( \langle B^{2}\rangle_{w}) $.~In this way it is possible to extract the real
parts of the second order weak values. Second order weak values for operators $A$ or $B$ can also be addressed separately, by considering independent Hamiltonians (i.e., taking either $B \equiv 0$, or $A \equiv 0$, respectively), as explained below. \\

We now consider in detail the case of a single observable weakly coupled to the measuring device (i.e., $B \equiv 0$). Eq. (\ref{Eq6}) shows that for $l>0$, an interaction that only couples the observable $A$ to the pointer can show a
nonzero expectation value for $\langle XY \rangle$, while  for Gaussian pointer states ($l=0$)
this is not the case. Specifically, for pointer states endowed with OAM, we obtain:

\begin{equation}
\label{Eq70} \langle XY \rangle= l \frac{\left(g_A t\right)^2}{2}
\Im \left(\langle A^2 \rangle_{w} \right).  
\end{equation}

Eq. (5) shows that it is possible to measure a value for $\langle XY \rangle
\ne 0$, even when the weak interaction does not couple the $y$-dimension of the pointer with the observable $A$ directly. This is
due to the fact that the pointer state given in Eq. (\ref{Eq5}) is
not factorable in the $x$- and $y$-dimensions, so the coupling of
the observable $A$ with the $x$-dimension of the pointer also
affects its $y$-dimension.~Eq.~(\ref{Eq70}) is quite significant as it readily shows that by
employing OAM pointer states, and by measuring the corresponding
two-dimensional pointer shift $\langle XY \rangle$, we can obtain an
imaginary weak value. 
We note that the imaginary part of the weak value 
provides information about the instantaneous rate of change in the probability due to the measurement process (i.e., back action mechanism) \cite{AS,Dressel}.~Such rate of change is usually linked to the pointers conjugate variable.~In our case, the non-factorability of the two-dimensional pointer spatial distribution produces a similar effect in the spatial domain, which can, in turn, be exploited to amplify the spatial shift, in a situation where Gaussian states can not provide an equivalent enhancement factor.  \\

Finally, we present a simple example which, never the less, fully demonstrates the advantages of employing OAM pointer states both in terms of signal amplification and information retrieval.  We consider a specific configuration which can be
experimentally implemented. This makes use of the ideas described
above, and will help to clearly reveal the advantages of employing
an OAM pointer state versus its 
Gaussian counterpart.~Consider a specific observable given by a
spin-1/2 matrix $A$, or the polarization degrees of freedom of the
radiation field. Such observable, which can be regarded generally
as a linear combination of Pauli matrices  \cite{Pauli}, has the
following form:
\begin{equation}
\label{Eq9} A=\left(
\begin{array}{cc}
9/5 & 2i/5 \\
-2i/5 & 6/5 \\
\end{array}
\right).
\end{equation}
To see its physical meaning in a weak measurement context, we can
calculate its eigenstates, $a$ and $b$, and the corresponding
eigenvalues, $\lambda_1$ and $\lambda_2$, respectively:
\begin{eqnarray}
\label{Eq10}
& & \lambda_1=1 \hspace{1cm} |a\rangle=1/\sqrt{5}\, \left( |H \rangle + 2i|V\rangle \right)\nonumber \\
& & \lambda_2=2 \hspace{1cm }|b\rangle=1/\sqrt{5}\, \left(
2i|H\rangle+|V\rangle \right).
\end{eqnarray}
Therefore, an interaction Hamiltonian of the form $H=g_A AP_x$
($B \equiv 0$), describes a process where for an arbitrary input
polarization, the $|a\rangle$ component is shifted by an amount
$\Delta=g_{A}t$, while the $|b\rangle$ component is shifted
by $2\Delta$. The initial state of the system is $|i\rangle
=|H\rangle$, and the initial shape of the pointer is
characterized by the distribution $\phi(x,y)$ given by Eq.
(\ref{Eq5}).~The initial state of the system and pointer
is factorable, and is given by $|\psi^{in}\rangle= |H\rangle \otimes \int dx dy
\phi(x,y) |x\rangle | y\rangle$.\\

The weak coupling interaction entangles the states of the system
and pointer such that the resulting output state can be written as:

\begin{equation}
\label{Eq12} |\psi^{out}\rangle=\int dx dy \left[ \phi(x-\Delta,y)|a\rangle-2i\phi(x-2\Delta,y) |b\rangle  \right] |x, y\rangle.
\end{equation}


The output state is projected into a nearly orthogonal state for the system 
$|f\rangle =\sin \epsilon |H\rangle+\cos \epsilon |V\rangle$, so
that $\langle f|i\rangle=\sin \epsilon$, with $\epsilon$ small.
When considering the OAM pointer distribution given in Eq.
(\ref{Eq5}), the expectation value of the pointer shift, up to
second order in $\Delta$, is given by:

\begin{equation}
\label{Eq14} \label{mean4} \langle XY \rangle=-\frac{3l}{5}\frac{
\Delta^2}{\tan \epsilon} .
\end{equation}

Note that when using a Gaussian
pointer state (here labeled with the subscript $\mathrm{G}$), we obtain $\langle XY \rangle_{\mathrm{G}}=0$, and the
usual one-dimensional mean value $\langle X \rangle_{\mathrm{G}}$ is:

\begin{eqnarray}
\label{mean2} \langle X \rangle_{\mathrm{G}}=\frac{9}{5} \Delta.
\end{eqnarray}

We also compare the first order weak value, given by:

\begin{equation}
\label{wv1}
\langle A \rangle_w=\frac{9}{5}-i\frac{2}{5}\frac{1}{\tan \epsilon},
\end{equation}

~with the higher order weak value, given by:
\begin{equation}
\label{wv2}
\langle A^2 \rangle_w=\frac{17}{5}-i\frac{6}{5}\frac{1}{\tan\epsilon}.
\end{equation}

Two points should be highlighted in this example. First, OAM pointer states allow retrieval of the imaginary part of the weak
value $\langle A^2 \rangle_w$, a feature which is not accessible with
Gaussian pointer states. Second, Eq. (\ref{mean2}) shows that
there is no weak amplification in the position shift  for a Gaussian pointer state. There is however amplification in the momentum shift $\langle
P_x\rangle_{\mathrm{G}}$, which is related to the imaginary part of the weak value (Eq. \ref{wv1}). On the other hand, the higher order weak value (Eq. \ref{wv2}) shows enhancement for small $\epsilon$.~This, as shown in Eq. (\ref{Eq70}), allows for weak amplification of the two-dimensional
position shift $\langle XY \rangle$.~We note that an important requirement in order to obtain such OAM enhancement is a non-vanishing  imaginary part for the second order weak value  (i.e., $\Im (\langle A^2 \rangle_{w}) \ne 0$).~This can, in turn, be interpreted as a specific form of ellipticity for the eigenvectors of the system observable $A$ \cite{Example}.\\

To conclude, we presented a novel scheme for 
weak measurements which relies on the use of pointer states
endowed with orbital angular momentum (OAM). We have shown that
such higher-dimensional weak measurements can provide access to
higher order weak moments of single-particle operators.~In
particular, for pointer states containing OAM with winding number
$l=\pm 1$, it is possible to measure the weak value of the square
of single-particle observables.

We have also shown that by considering a single-particle operator
and a two-dimensional OAM pointer state, it is possible to measure
an imaginary weak value via a two-dimensional pointer
position shift. This result is unexpected, since it is usually
believed that imaginary weak values are only accessible through the expectation value of the pointers conjugate
variable. In addition,
we have demonstrated, by means of an example, that the use
of OAM pointer states can provide weak amplification in configurations
where the use of Gaussian pointer states cannot, thus allowing for a much wider range of applicability. For
instance, one can engineer OAM states with higher winding numbers,
or superpositions of OAM states, to obtain the sought-after
higher order weak values that show the required enhancement. 

The results presented here open the door to a number of novel fundamental and technological applications  \cite{Briegel, Bell, Granata, Hawking}.
Furthermore, we emphasize that the use of pointer states with OAM
is not restricted to radiation fields, and could also be
envisioned in the context of Bose-Einstein condensates,
where the coherent transfer of OAM  of photons to matter
can be used to create atomic vortex states \cite{andersen2006}.\\

This work was supported by the Government of Spain (Project
FIS2010-14831) and project FET-Open 255914 (PHORBITECH). This work
has also been supported by Fundacio Privada Cellex Barcelona. GP
acknowledges financial support from Marie Curie International
Incoming Fellowship COFUND.\\

{}


\begin{thebibliography}{}
\bibitem{Example} A particular example where our scheme would not retrieve any advantage over Gaussian pointer states, is given for system observables satisfying $A^2=1$, such as the Pauli matrices. We note, however, that the maximum weak-value amplification factor that can be obtained  in this particular case, employing Gaussian probe states, is also bounded \cite{Upperbound}.
\end{thebibliography}
\end{document}